\documentclass[10pt,twocolumn]{article}

\usepackage[margin=0.72in]{geometry}
\usepackage{cite}
\usepackage{amsmath,amssymb,amsfonts}
\usepackage{algorithmic}
\usepackage{graphicx}
\usepackage{textcomp}
\usepackage{subcaption}
\usepackage{siunitx}
\usepackage{multirow}

\title{SEU Hardened Multi-Context Spintronic FPGA}
\author{%
    Abdolah Amirany$^{1}$, HamidReza Imani$^{2}$,\\
    Samuel Farid$^{2}$, and Tarek El-Ghazawi$^{2}$\\[0.65em]
    \small $^{1}$Herbert Wertheim College of Engineering, University of Florida,\\
    \small Gainesville, FL 32611, USA\\
    \small $^{2}$Department of Electrical and Computer Engineering, The George Washington University,\\
    \small Washington, DC 20052, USA\\[0.55em]
    \small \texttt{a\_amirany@ieee.org} \quad \texttt{hamidreza@gwu.edu}\\
    \small \texttt{sfarid@gwu.edu} \quad \texttt{tarek@gwu.edu}
}
\date{}

\begin{document}
\maketitle

\begin{abstract}
Due to susceptibility to radiation-induced soft errors and resource constraints implementing large and complex designs on field programmable gate arrays (FPGAs) in harsh environments is challenging. Therefore, to mitigate these issues, this paper proposes an innovative approach through the introduction of a single event upset (SEU)-hardened spintronic FPGA architecture that incorporates multi-context capabilities. Leveraging the non-volatile properties of magnetic tunnel junction (MTJ), the proposed design can store two/four distinct configurations internally, eliminating the dependence on external non-volatile memory which significantly boosts the system performance. Additionally, through post-layout simulation, we demonstrate our designs’ immunity to SEU, which enhances its reliability and operational stability in environments susceptible to radiation-induced soft errors. The efficiency of the proposed design was also evaluated using the ROSSETA benchmark, which serves as a representative model for real-world FPGA applications.
\end{abstract}


\section{Introduction}
\label{sec:introduction}
Complex programmable logic devices (CPLDs) and field-programmable gate arrays (FPGAs) that utilize static random-access memories (SRAM) are extensively employed in circuit design due to their outstanding characteristics~\cite{zhao2009sttmram,rajaei2018hybridsrammram}. These characteristics include superior performance and the ability to be reconfigured infinitely~\cite{rajaei2018hybridsrammram}. However, despite these advantages and the considerable hardware capabilities these chips offer, they come with their own limitations. Among the most important limitations are volatility, resource limitation, and single event upset (SEU) vulnerability~\cite{charitopoulos2015taskscheduling,amirany2020retentionlatch, suvizi2025auto}.

Firstly, SRAM is a volatile memory, meaning that it loses the information stored in it as soon as the power supply is cut off~\cite{rajaei2018hybridsrammram}. This characteristic necessitates the use of an external, non-volatile memory to permanently store FPGA configurations. Typically, this role is filled by external flash-based memories~\cite{rajaei2018hybridsrammram,rajaei2016radiationhardened}. However, flash-based memories suffer from limited endurance and high-power consumption. Moreover, transferring the configuration data from external memory to FPGA is not efficient in terms of delay and power consumption~\cite{zhao2009sttmram,rajaei2018hybridsrammram, suvizi2025parallel}.

Secondly, although CPLDs and FPGAs are equipped with substantial hardware capabilities, they face limitations when it comes to executing very large and complex designs~\cite{charitopoulos2015taskscheduling,ramezani2020dynamicscheduling}. To overcome these challenges, designers often rely on using multiple chip configurations or consider task scheduling~\cite{scalera1998contextswitching,anderson2020roc,suvizi2024exploring}. However, employing multiple chips results in the use of several distinct chips, leading to a significant increase in costs~\cite{zhao2009sttmram,elghazawi2008promise}. On the other hand, task scheduling necessitates continuous access to external memory, which can considerably degrade the overall system performance~\cite{amirany2021taskschedulable}.

Thirdly, SRAM-based FPGAs and CPLDs are susceptible to SEUs. SEUs are transient faults caused by high-energy particles striking the semiconductor material which result in unintended changes in the stored data~\cite{bakhtiary2023ternarysram}. This vulnerability is a significant concern specifically in environments with high levels of radiation, such as outer space or near nuclear reactors, where particle interference is more probable~\cite{glorieux2013dff}. The sensitivity of SRAM-based devices to SEUs poses challenges regarding their reliability. As a results, it is necessary to implement robust error detection and correction mechanisms to mitigate the risks associated with these transient faults which severely worsen the resource limitation issues~\cite{rajaei2016radiationhardened}.

Magnetic tunnel junction (MTJ) based memories present a solution to the volatility issue inherent in SRAM-based chips~\cite{bahmanabadi2022quaternarymemory,wang2020nonvolatilesram}. MTJs are spintronic devices that store information using magnetic vectors, allowing data to be preserved even when power is not supplied~\cite{wang2015perpendicular,amirany2021mtmrsnqm}. Furthermore, MTJs offer compatibility with complementary metal–oxide semiconductor (CMOS) technology, an almost unlimited lifespan, low static power consumption, and intrinsic immunity to radiation induced soft errors~\cite{raouf2023fulladder,grollier2020neuromorphic, adelkhani2025lpcr}.

Although MTJs are immune to SEU, the peripheral CMOS circuits needed for MTJ reading are SEU susceptible~\cite{rajaei2018hybridsrammram,rajaei2016radiationhardened}.  To address this issue and resource limitation of the FPGAs and CPLDs, this paper proposes an SEU-hardened multi-context spintronic FPGA design. The proposed design in this paper, without the need for any external memory, can store two/four completely unique configurations and can switch between these configurations seamlessly.

The rest of the paper is organized as follows. Section~\ref{sec:background} reviews the paper background including spintronic technology and SRAM-based FPGA architecture. Section~\ref{sec:previous_work} investigates the previous spintronic FPGA designs, their advantages, and disadvantages. The proposed SEU-hardened multi-context spintronic FPGA design is explained in Section~\ref{sec:proposed}. In section~\ref{sec:results}, the simulation results are presented to show the performance and validate the functionality of the proposed design. Finally, the paper is concluded in Section~\ref{sec:conclusion}.

\section{Background}
\label{sec:background}
\subsection{Spintronic}
Spintronics, short for spin electronics, diverges from traditional electronics by not only using the electron's charge but also its spin and the resulting magnetic moment for data processing and storage~\cite{stephan2020spinhall}. 

\begin{figure}
    \centering
    \includegraphics[width=1\linewidth]{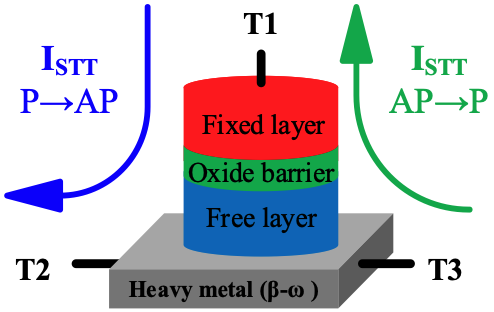}
    \caption{Conceptual structure of an MTJ}
    \label{fig:mtj_structure}
\end{figure}

The MTJ is the fundamental component of spintronic circuits, consisting of two ferromagnetic layers (A.K.A, the free and the reference layer) separated by an ultra-thin insulator, all mounted on a heavy metal strip~\cite{wang2015perpendicular}. Figure~\ref{fig:mtj_structure} shows the conceptual structure of an MTJ~\cite{wang2015perpendicular}.

Experimental observations have shown that when the insulator layer is sufficiently thin, electrons can tunnel from one ferromagnetic layer to the other, generating an electric current~\cite{wang2015perpendicular}. This tunneling is influenced by the relative orientation of the magnetic vectors of the ferromagnetic layers, leading to two distinct operational states for an MTJ:
\begin{enumerate}
    \item Parallel state: The magnetic vectors of the ferromagnetic layers align in the same direction, facilitating electron tunneling and resulting in low electrical resistance ($R_P$).

    \item Antiparallel state: The magnetic vectors are oriented in opposite directions, which prevents electron tunneling and increases electrical resistance ($R_{AP}$).
\end{enumerate}

This behavior enables the MTJ to function as a switch with two non-volatile modes~\cite{jamshidi2018puremagnetic}. The difference in resistance between these states defines the tunnel magnetoresistance ratio (TMR)~\cite{ywang2014compactmodel}. TMR ratio is expressed by Eq.~\ref{eqn:tmr}.

\begin{equation}
\label{eqn:tmr}
\mathrm{TMR} = \frac{R_{\mathrm{AP}} - R_{\mathrm{P}}}{R_{\mathrm{P}}} \times 100\%
\end{equation}

TMR ratio is a critical parameter for the MTJ's readability and read stability and a higher TMR ratio indicates better performance~\cite{jamshidi2018puremagnetic,amirany2021stochasticneuron}.

\subsection{SRAM-Based FPGAs and CPLDs}
\begin{figure}
    \centering
    \includegraphics[width=1\linewidth]{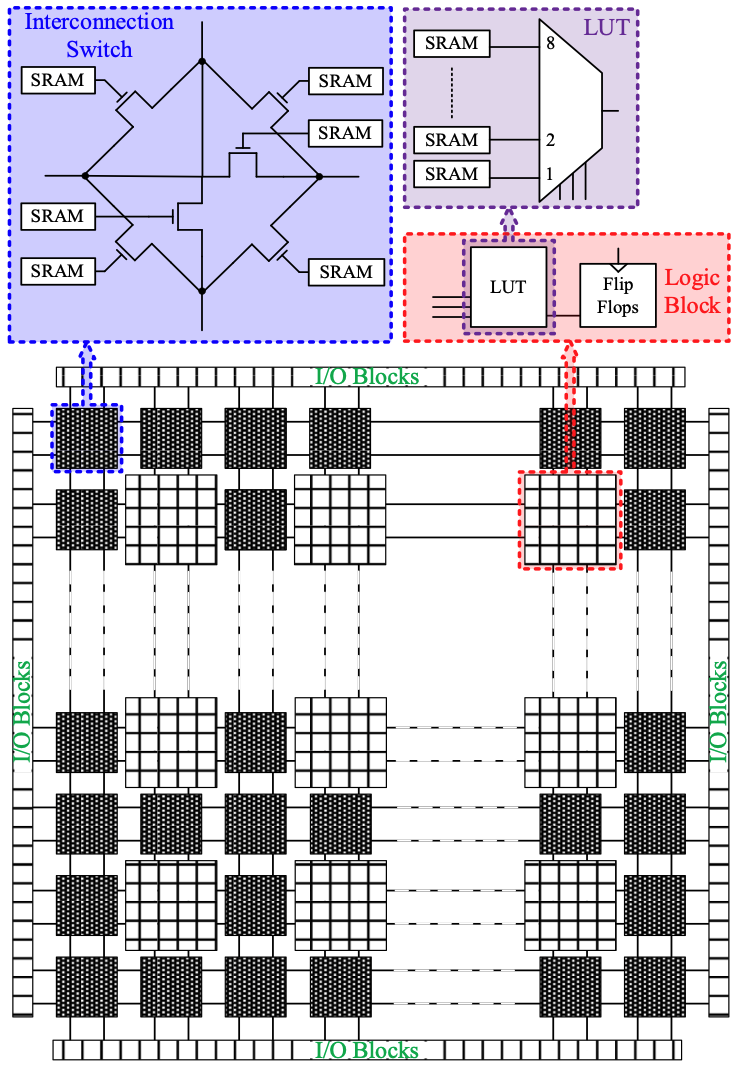}
    \caption{SRAM-based FPGA/CPLD structure}
    \label{fig:structure}
\end{figure}

Figure~\ref{fig:structure} illustrates the structure of an SRAM-based FPGA/CPLD~\cite{rajaei2018hybridsrammram}. This schematic breaks down an FPGA/CPLD into three fundamental components~\cite{amirany2021taskschedulable}: 
\begin{enumerate}
    \item I/O blocks
    \item Logic blocks
    \item Interconnection switches 
\end{enumerate}

I/O blocks are responsible for data exchange between the FPGA and external systems. Logic blocks execute logic operations, utilizing look-up tables (LUTs) for storing information and function implementation. Lastly, interconnection switches are responsible for linking different sections of the FPGA~\cite{zhao2009sttmram,rajaei2018hybridsrammram,charitopoulos2015taskscheduling}. 

Figure~\ref{fig:structure} also highlights two out of the three main parts of the FPGA, interconnection switches and logic blocks, which are based on SRAM technology~\cite{amirany2021taskschedulable}. SRAM cells within interconnection switches dictate the routing paths, whereas in logic blocks, SRAMs store data and implement logical functions~\cite{zhao2009sttmram,rajaei2018hybridsrammram}. 

As stated in the introduction, the reliance on SRAM introduces a significant limitation due to its volatile nature. This volatility necessitates the use of an external non-volatile memory to preserve the FPGA's configuration data. This prevents the functionality to be compromised during power outage~\cite{zhao2009sttmram,rajaei2018hybridsrammram}.

\section{Previous Work}
\label{sec:previous_work}

Previous efforts in the field of spintronic FPGAs/CPLDs can be categorized into three groups, which are discussed below.

\subsection{Spintronic FPGAs/CPLDs}
This type of spintronic FPGAs/CPLDs are of the most common designs. In this type, MTJs are used as the non-volatile memory for storing configurations. Therefore, in addition to non-volatility, this type also benefits from low static power consumption and the infinite endurance of the MTJs~\cite{zhao2012crosspoint}. Despite the aforementioned advantages, this type is susceptible to SEU-induced soft errors and also has limited resources. These disadvantages restrict the application of this type of CPLDs/ FPGAs, especially when implementing large and complex designs or for use in harsh environments.

\subsection{SEU-Hardened Spintronic FPGAs/CPLDs}
This type of SEU-hardened spintronic FPGAs/CPLDs represent an advanced category within the field. In this type, similar to the first type, MTJs serve as a non-volatile memory element to store configurations. Therefore, in addition to inherent non-volatility, low static power consumption, and the infinite endurance characteristics of MTJs, they provide SEU immunity~\cite{rajaei2018hybridsrammram,rajaei2016radiationhardened}. Furthermore, their enhanced design techniques increase their resilience against SEU-induced soft errors. Despite these improvements, the SEU-hardened spintronic FPGAs/CPLDs may still face challenges related to limited resources, which can constrain their utility in implementing large and complex designs.

~\subsection{Task-Schedulable Spintronic FPGAs/CPLDs}
This category of Task-Schedulable Spintronic FPGAs/CPLDs stands out for its unique feature of task scheduling capability. This type also utilizes MTJs for storing configurations in a non-volatile manner~\cite{amirany2021taskschedulable}. The key advantage of this type is task scheduling capability, which makes it a suitable choice for implementations of large and complex designs. However, unlike their SEU-hardened counterparts, this type does not incorporate specific design enhancements for mitigating SEU, rendering them vulnerable to SEU-induced soft errors. 

\section{Proposed SEU-Hardened Multi-Context Spintronic FPGA Structure}
\label{sec:proposed}

In this section, first we introduce our SEU-hardened magnetic random-access memory (MRAM) that incorporates multi-context functionality. Following this, we detail the design of our SEU-hardened spintronic LUT that is also equipped with multi-context capability. We conclude this section by outlining the design framework for the proposed SEU-hardened multi-context spintronic FPGA that integrates both proposed SEU-hardened multi-context MRAM and SEU-hardened multi-context LUT.

\subsection{Proposed SEU-Hardened Multi-Context MRAM}
\begin{figure}
    \centering
    \includegraphics[width=1\linewidth]{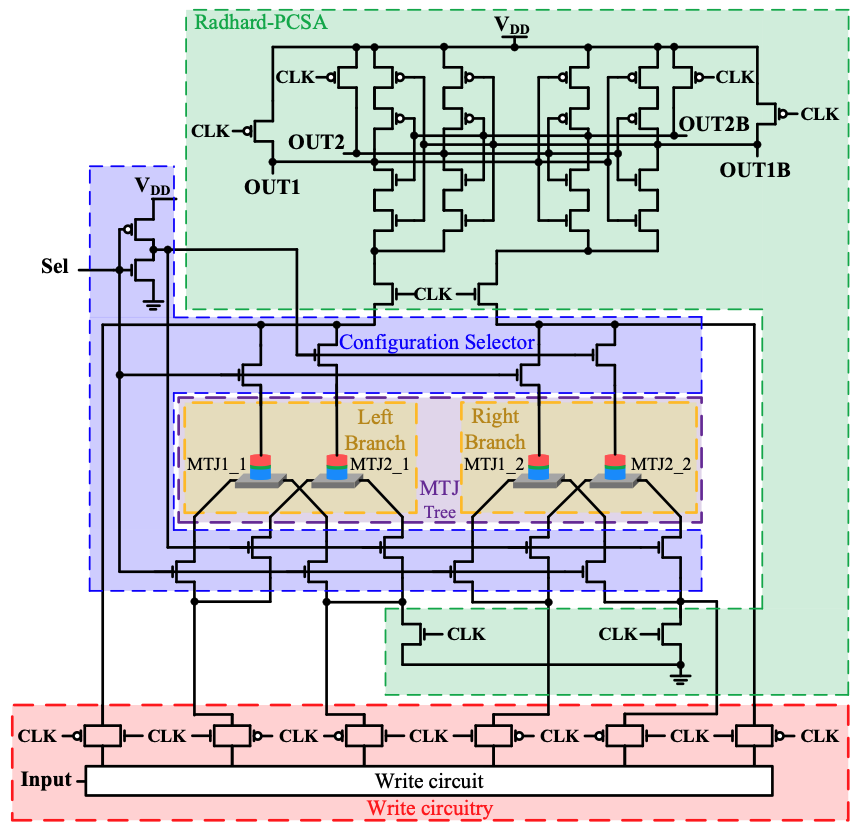}
    \caption{Proposed SEU-hardened multi-context MRAM circuit}
    \label{fig:proposed_circuit}
\end{figure}

Figure~\ref{fig:proposed_circuit} illustrates the schematic of the proposed SEU-hardened multi-context MRAM, which is composed of four key components:
\begin{enumerate}
    \item SEU-Hardened Pre-Charge Sense Amplifier (PCSA): This component is fundamental in reading the states of MTJs and generating the corresponding outputs (OUT1, OUT1B, OUT2, and OUT2B). The proposed SEU-hardened PCSA is comprised of four C-elements, which are crucial components that lend the design its distinctive characteristics of SEU immunity. C-elements, as described in the literature, are three-terminal devices that typically function as inverters when their inputs are of the same logical state. Conversely, when the inputs exhibit differing logical states, the output of a C-element transitions to a high impedance state, effectively becoming isolated from the circuit. Should any of the nodes, OUT1, OUT1B, OUT2, and OUT2B, be impacted by energetic particles, the outputs of the C-elements connected to these nodes will transition to a high impedance state. This response isolates these nodes from any subsequent variations, thereby preventing the propagation of soft errors throughout the circuit. This mechanism ensures that the structure maintains its operational integrity even in the presence of potentially disruptive energetic particle impacts. Its functionality is centered on the capacitive nodes' discharge process. When the clock signal is in the '1' state, the MTJ branch with the lower resistance (parallel MTJ) discharges its associated nodes faster, leading to its corresponding outputs becoming '0', while the other outputs remain '1'.

    \item Configuration Selector: This segment is instrumental in enabling multi-context switching within the circuit. Based on the selector signal (Sel), it connects one of the two MTJ branches to the PCSA circuit and the write circuitry, allowing for the storage and retrieval of two distinct configurations. Remarkably, this functionality eliminates the need for external memory to access these configurations.

    \item MTJs Tree: Comprising of four MTJs divided into two branches (left branch and right branch), this component is fundamental to the storage of information. The left branch consists of $MTJ1\_1$ and $MTJ2\_1$, while the right branch is made up of $MTJ1\_2$ and $MTJ2\_2$. Each pair of MTJs within these branches is tasked with storing a single bit of information.

    \item Write Circuitry: When the clock signal (CLK) is '0' this component is tasked with configuring the MTJs according to the input signal (Input).
\end{enumerate}

\begin{figure*}
    \centering
    \includegraphics[width=1\linewidth]{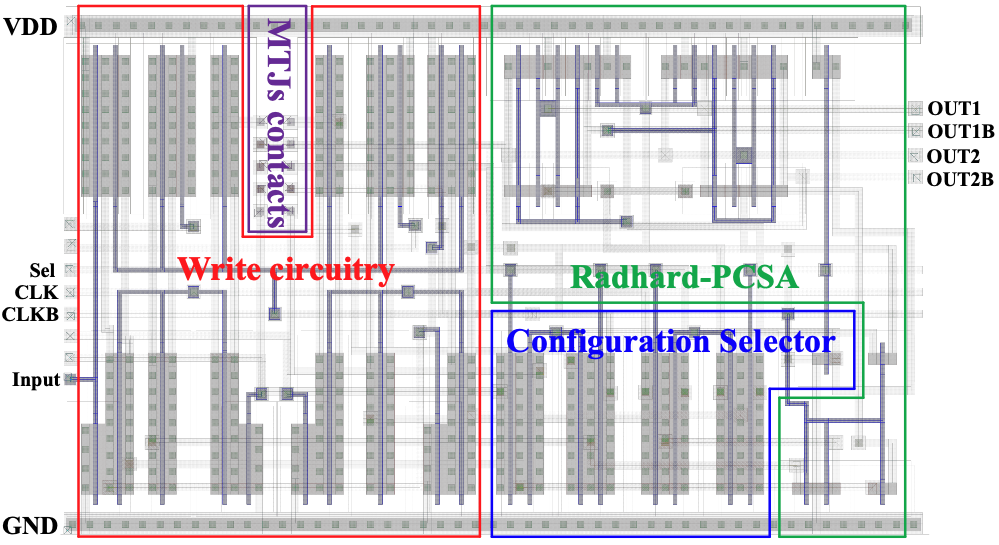}
    \caption{Layout of the proposed SEU hardened multi-context MRAM circuit}
    \label{fig:layout}
\end{figure*}

Figure~\ref{fig:layout} shows the layout view of our proposed SEU-hardened task-schedulable MRAM circuit designed using 40nm CMOS technology. It is worth mentioning that the proposed circuit occupies a 36.66$\mu\text{m}^2$ area.

\subsection{Proposed SEU-Hardened Multi-Context LUT}
The proposed SEU-hardened multi-context LUT circuit shares the same structure with proposed SEU-hardened multi-context MRAM circuit, with the primary distinction in the MTJs tree and configuration selector. The MTJ tree in this circuit consists of two sets of MTJ trees, each containing 16 MTJs. These trees are accompanied by 28 transistors functioning as a multiplexer, with each tree dedicated to storing one configuration.

\begin{figure}
    \centering
    \includegraphics[width=1\linewidth]{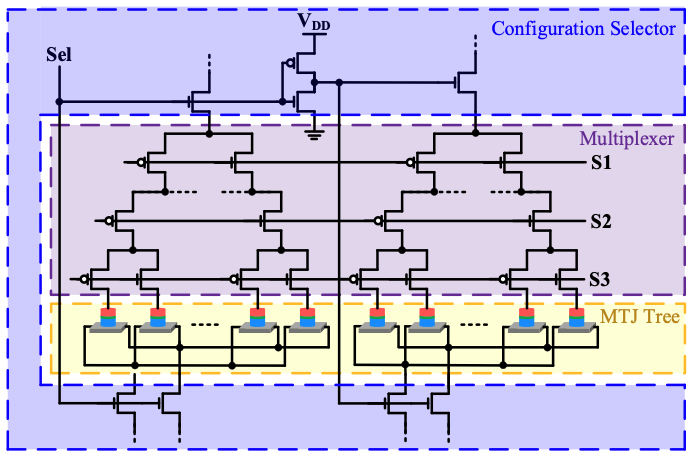}
    \caption{Left branch of the proposed SEU-hardened multi-context LUT circuit}
    \label{fig:left_branch}
\end{figure}

Figure~\ref{fig:left_branch} illustrates the left branch of MTJs which is required for implementing the proposed SEU-hardened multi-context LUT. In this figure, transistors gated by signals S1, S2, and S3 operate as multiplexers. Depending on the states of these signals, an MTJ is connected to the PCSA circuit, allowing the data stored within the selected MTJ to be transferred to the output. 

For the SEU-hardened multi-context LUT design, the left branch depicted in Fig.~\ref{fig:proposed_circuit} is substituted with the configuration shown in Fig.~\ref{fig:left_branch}. The right branch mirrors the left, with the exception that the data stored in the MTJs of the right branch is the complement of the data stored in the left branch MTJs. This design ensures that the SEU-hardened multi-context LUT can effectively store and select between two distinct configurations.

\subsection{Proposed SEU-Hardened Multi-Context FPGA Architecture}
The architecture of the proposed SEU-hardened multi-context non-volatile FPGA is modeled closely on the structure depicted in Fig.~\ref{fig:mtj_structure}, with some modifications. These modifications include the substitution of traditional SRAM cells within the interconnection switches with the proposed SEU-hardened multi-context MRAM and substitution of LUTs within the logic blocks with the proposed SEU-hardened multi-context LUT. It is also noteworthy that by adding one more stage to the configuration selector and more MTJs to the MTJs tree the proposed circuits can be redesigned in a way that can store 4 configurations.

\section{Simulation Results}
\label{sec:results}

\begin{table}[t]
\caption{Critical Parameters of the MTJs and Transistors}
\label{tab:mtj_params}
\centering
\resizebox{\columnwidth}{!}{%
\begin{tabular}{|l|c|}
\hline
\textbf{Parameter} & \textbf{Value} \\
\hline

\multicolumn{2}{|c|}{\textbf{MTJ}} \\
\hline

Shape & Round \\
MTJ diameter & 55 nm \\
Barrier thickness & 0.9 nm \\
Free layer thickness & 0.9 nm \\
Fixed layer thickness & 3 nm \\
TMR under zero bias voltage ($\mathrm{TMR}_0$) & 200\% \\
TMR under 1 V bias voltage & 180\% \\
Resistance area product & $10\,\Omega\cdot\mu\mathrm{m}^2$ \\
Perpendicular magnetic anisotropy & $10 \times 10^4$ A/m \\
Free layer saturation field & $11 \times 10^5$ A/m \\
MTJ resistance & $10 \sim 30$ k$\Omega$ \\
$\beta$-W strip volume & $60\,\mathrm{nm} \times 100\,\mathrm{nm} \times 3\,\mathrm{nm}$ \\
$\beta$-W strip resistivity & $2\,\Omega\cdot\mathrm{m}$ \\
Characteristic time & 2 ns \\

\hline
\multicolumn{2}{|c|}{\textbf{Transistors}} \\
\hline

Gate length & 40 nm \\
Gate width & See Fig.~\ref{fig:proposed_circuit} \\
Gate oxide thickness & 2 nm \\
Supply voltage & 1.0 V \\

\hline
\end{tabular}
}
\end{table}

Hspice post-layout simulations were conducted to validate functionality and performance of the proposed design. In all simulations, 40 nm CMOS technology and MTJ model proposed in~\cite{wang2015perpendicular} were employed. In the simulations conducted, the supply voltage (VDD) is set to 1V. A pulse signal is applied as the clock (CLK) signal, with a 50$\%$ duty cycle and an 8ns period. The parameters for MTJs and transistors are provided in Table~\ref{tab:mtj_params}. This setup ensures a standardized testing environment across simulations. Also, For the purpose of SEU injection in simulations, a transient double exponential current source model, as proposed in~\cite{rockett1988seulatch,jain2005seureliability}, is employed. This model is defined by Eq.~\ref{eqn:model}.

\begin{equation}
\label{eqn:model}
I_{\mathrm{inj}}(t) = \frac{Q_{\mathrm{inj}}}{\tau_1 - \tau_2}
\left( e^{-t/\tau_1} - e^{-t/\tau_2} \right)
\end{equation}

where $Q_{inj}$ represents the quantity of charge injected into the node affected by energetic particles. Additionally, $\tau1$ and $\tau2$ are material-dependent time constants. This model has been extensively utilized by researchers in various studies due to its high accuracy and producing results that closely align with experimental data~\cite{bakhtiary2023ternarysram}. In the SEU injection simulations, the charge amount $Q_{inj}$ is varied from -20 femto coulombs (fC) to 20fC, with increments of 5fC. This range is considered based on the understanding that particles carrying a charge in the femto coulomb range, approximately 0.1fC to 5fC, are capable of causing SEUs. Therefore, testing with a particle charge of 20fC provides a comprehensive validation of the SEU tolerance capabilities of the proposed circuit.

It is worth noting that all designs considered in this section~\cite{rajaei2016radiationhardened,rajaei2018hybridsrammram,amirany2021taskschedulable,zhao2012crosspoint,wu2016temperatureimpact} were simulated under identical conditions to ensure a fair and accurate comparison. We also use equivalent unit size transistors (UST) for area comparison~\cite{amirany2020trng}.

\subsection{Functional Simulation and SEU Injection}

\begin{figure}
    \centering
    \includegraphics[width=1\linewidth]{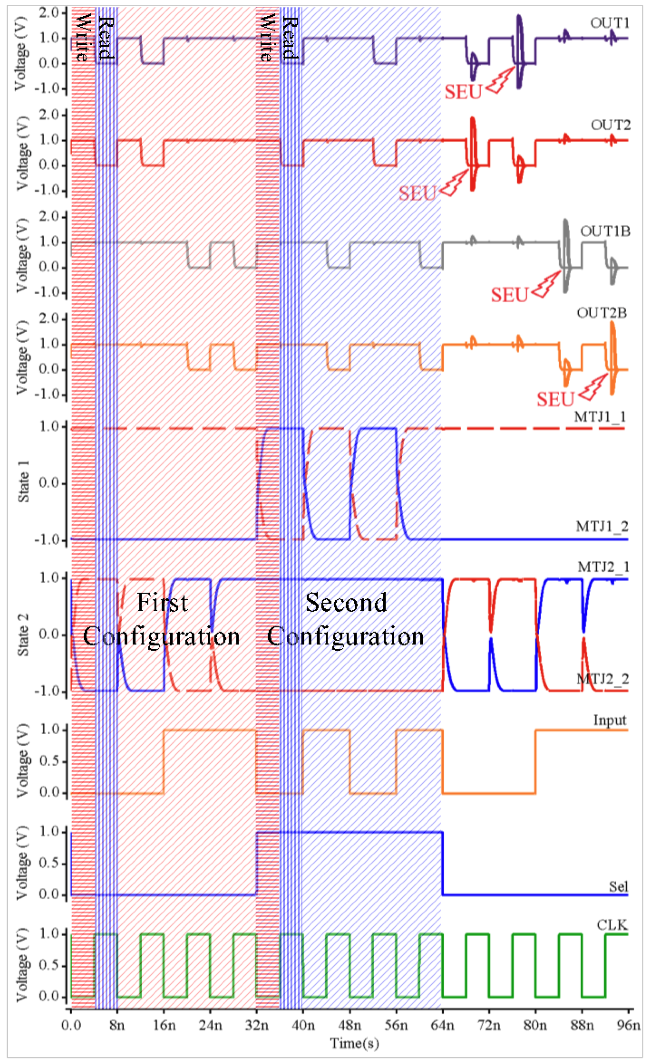}
    \caption{Transient simulation results for the proposed SEU hardened Multi-context MRAM}
    \label{fig:sim_results}
\end{figure}

Figure~\ref{fig:sim_results} illustrates the transient simulation results for the proposed SEU-hardened multi-context MRAM, showing its capability to store and retrieve two distinct configurations. Initially, with the Sel signal set to '0', $MTJ1\_1$ and $MTJ1\_2$ are connected for read and write operations. However, when the Sel signal changes from ‘0’ to '1', the operations are directed through $MTJ2\_1$ and $MTJ2\_2$. This enables the circuit to manage two configurations using a single Sel signal, thereby avoiding the need for external memory which significantly boosts the FPGA performance for multi-context applications.

Another key factor in MTJ-based circuits is the retention time, which refers to the duration for which the MTJ can retain its stored value. The retention time for an MTJ is given by Eq.~\ref{eqn:t_ret}~\cite{ywang2014compactmodel}.

\begin{equation}
\label{eqn:t_ret}
T_{\mathrm{ret}} = t_0 \times \exp\left(
\frac{0.5\,\mu_0 \times H_k \times M_s \times V}{k_B \times T}
\right)
\end{equation}

where $t_0$ is the characteristic time, $\mu_0$ is the vacuum permeability, $H_k$ is the perpendicular magnetic anisotropy, $M_s$ is the Free layer saturation field, $V$ is the volume of the free layer, $k_B$ is the Boltzmann constant, and $T$ is the absolute temperature. Using the parameter values from Table~\ref{tab:mtj_params}, the calculated retention time is 13.9 years, which is sufficient for reliable operations in FPGA configuration memory. 

\subsection{Compratative Analyses}

\begin{table*}[]
\caption{Performance Analysis of the Proposed SEU-Hardened Multi-Context Spintronic FPGA}
\label{tab:fpga_perf}
\centering

\resizebox{\textwidth}{!}{%
\begin{tabular}{|l|l|c|c|c|c|c|c|c|}
\hline
\multicolumn{2}{|c|}{Design} &
\begin{tabular}[c]{@{}c@{}}\\Proposed \\in \scriptsize~\cite{zhao2012crosspoint}\end{tabular} &
\begin{tabular}[c]{@{}c@{}}\\Proposed \\in \scriptsize~\cite{wu2016temperatureimpact}\end{tabular} &
\begin{tabular}[c]{@{}c@{}}\\Proposed \\in \scriptsize~\cite{rajaei2016radiationhardened}\end{tabular} &
\begin{tabular}[c]{@{}c@{}}\\Proposed \\in \scriptsize~\cite{rajaei2018hybridsrammram}\end{tabular} &
\begin{tabular}[c]{@{}c@{}}\\Proposed \\in \scriptsize~\cite{amirany2021taskschedulable}\end{tabular} &
\multicolumn{2}{c|}{Proposed in this paper} \\
\cline{8-9}
\multicolumn{2}{|c|}{} & & & & & & \scriptsize MC2 & \scriptsize MC4 \\
\hline

\multirow{2}{*}{Area$^*$}
& \# of UST & 170 & 100 & 135 & 100 & 105 & 130 & \textbf{85} \\
\cline{2-9}
& \# of MTJ & \textbf{2} & \textbf{2} & \textbf{2} & \textbf{2} & \textbf{2} & \textbf{2} & \textbf{2} \\
\hline

\multirow{2}{*}{Power$^*$}
& Average ($\mu$W) & 19.9 & 9.1 & 11.2 & 6.1 & 10.5 & 10.4 & \textbf{4.7} \\
\cline{2-9}
& Static (nW) & 0.7 & 7147.1 & 1.6 & 0.7 & \textbf{0.6} & 2.9 & 4.1 \\
\hline

\multirow{2}{*}{Read Delay (ps)}
& Average & 101.3 & 126.8 & 254.1 & 235.2 & \textbf{77.3} & 130.2 & 99.1 \\
\cline{2-9}
& Maximum & 104.8 & 158.6 & 270.3 & 240.1 & \textbf{77.6} & 130.8 & 103.3 \\
\hline

\multicolumn{2}{|l|}{Write Delay (ns)} & \textbf{1.03} & 3.29 & 5.21 & 2.01 & 1.22 & 1.42 & 1.68 \\
\hline

\multicolumn{2}{|l|}{PDP (fJ)} & 2.02 & 1.15 & 2.85 & 1.43 & 0.81 & 1.35 & \textbf{0.47} \\
\hline

\multicolumn{2}{|l|}{MTJ write access} & 1 & 1 & 1 & 1 & \textbf{0.5} & 0.5 & \textbf{0.25} \\
\hline

\multicolumn{2}{|l|}{Multi-context capability?} & No & No & No & No & \textbf{Yes} & \textbf{Yes} & \textbf{Yes} \\
\hline

\multicolumn{2}{|l|}{SEU Immunity?} & No & No & \textbf{Yes} & \textbf{Yes} & No & \textbf{Yes} & \textbf{Yes} \\
\hline

\multicolumn{9}{l}{$^*$Per configuration}
\end{tabular}
}
\end{table*}

As Table~\ref{tab:fpga_perf} shows, the proposed design exhibits significant improvements in power consumption compared to other SEU-hardened designs~\cite{rajaei2016radiationhardened,rajaei2018hybridsrammram}. Notably, our design reduces the average power consumption from 22$\%$ to 50$\%$ when compared against prior SEU-immune designs. Even when compared to the non-SEU-immune design of~\cite{amirany2021taskschedulable}, our design shows a considerable reduction in power usage while still providing SEU protection features and multi-context capability. This highlights our design’s capability to provide SEU immunity and multi context capabilities simultaneously while requiring lower amounts of power.

Regarding the area, the proposed design achieves a UST reduction of 15$\%$ to 35$\%$ compared to SEU-immune designs proposed in~\cite{rajaei2016radiationhardened,rajaei2018hybridsrammram}. Also, as Table~\ref{tab:fpga_perf} shows, the proposed design achieves a power-delay product (PDP) improvement of 40$\%$ to 60$\%$ compared to all considered designs in Table~\ref{tab:fpga_perf}. This demonstrates that our design achieves a more optimal position on the Pareto frontier, balancing energy consumption and performance more effectively.

\subsection{Benchmarking}

\begin{table}[]
\caption{ROSSETA Benchmark Specification}
\label{tab:rosetta}
\centering

\setlength{\tabcolsep}{4pt}
\renewcommand{\arraystretch}{1.1}

\begin{tabular}{|l|c|c|}
\hline
\textbf{Test suits} & \textbf{LUTs} & \textbf{FFs} \\
\hline
3D rendering (3DR) & 8893 & 12471 \\
Digit Recognition (DR) & 41238 & 26468 \\
Spam Filtering (SF) & 12678 & 22134 \\
Optical Flow (OF) & 42878 & 61078 \\
Binarized Neural Network (BNN) & 46899 & 46760 \\
Face Detection (FD) & 62688 & 83804 \\
\hline
\end{tabular}
\end{table}

We use ROSETTA~\cite{zhou2018rosetta} benchmark suit to evaluate the performance of the proposed SEU-hardened multi-context spintronic FPGA. The ROSSETA benchmark suite provides standardized real-world applications, such as machine learning and video processing. Table~\ref{tab:rosetta} provides a detailed account of the number of LUTs and FFs of this benchmark and its respective subsections. 

\begin{figure}
    \centering
    \includegraphics[width=1\linewidth]{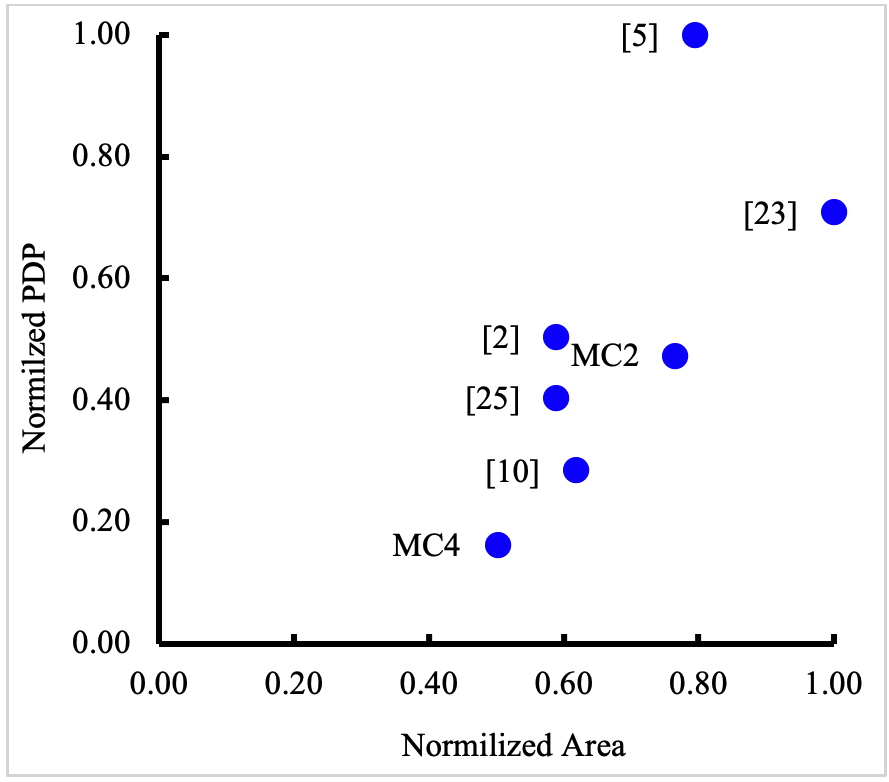}
    \caption{Results of ROSSETA~\cite{zhou2018rosetta} benchmark}
    \label{fig:results}
\end{figure}

Figure~\ref{fig:results} shows the average results of the ROSSETA benchmark, comparing various designs in terms of normalized PDP and normalized area. As shown in this figure, our proposed designs offer significant improvements over other designs by being able to store two/four configurations without incurring substantial area overhead. The proposed 4-configurations (MC4) design achieves the lowest PDP and area, showcasing its efficiency in minimizing both power consumption and area overhead. The proposed 2-configurations (MC2) design also performs well, with a lower PDP than other SEU-hardened designs while maintaining a relatively compact area. These results highlight the effectiveness of our SEU-tolerant, multi-context spintronic FPGA in real-world applications covered by the ROSSETA benchmark, offering superior performance in energy consumption and area efficiency compared to existing designs while bringing SEU immunity and multi-context capability.

\section{Conclusion}
\label{sec:conclusion}
In this paper, we introduced an SEU-hardened multi-context spintronic FPGA architecture. By integrating MTJ technology as a non-volatile memory storage, this design offers a dual advantage of storing multiple configurations internally while ensuring resilience against SEU without significant area overhead or power consumption. This approach effectively eliminates the need for external non-volatile memory which reduces power consumption and enhances the system's performance and reliability. Post-layout simulation shown that, proposed architecture showcases a notable reduction in power consumption, achieving a minimum of 23$\%$ lower average power and 14$\%$ lower static power consumption compared to existing designs. The proposed SEU-hardened multi-context spintronic FPGA not only meets the challenges posed by large and complex design requirements but also can be used in environments susceptible to radiation-induced errors.

\bibliographystyle{IEEEtran}
\bibliography{refs}

@article{zhao2009sttmram,
  title={Spin transfer torque (STT)-MRAM--based runtime reconfiguration FPGA circuit},
  author={Zhao, Weisheng and Belhaire, Eric and Chappert, Claude and Mazoyer, Pascale},
  journal={ACM Transactions on Embedded Computing Systems (TECS)},
  volume={9},
  number={2},
  pages={1--16},
  year={2009},
  publisher={ACM New York, NY, USA}
}

@article{rajaei2018hybridsrammram,
  title={Reliable, high-performance, and nonvolatile hybrid SRAM/MRAM-based structures for reconfigurable nanoscale logic devices},
  author={Rajaei, Ramin and Amirany, Abdolah},
  journal={Journal of Nanoelectronics and Optoelectronics},
  volume={13},
  number={9},
  pages={1271--1283},
  year={2018},
  publisher={American Scientific Publishers}
}

@inproceedings{charitopoulos2015taskscheduling,
  title={Hardware task scheduling for partially reconfigurable FPGAs},
  author={Charitopoulos, George and Koidis, Iosif and Papadimitriou, Kyprianos and Pnevmatikatos, Dionisios},
  booktitle={International symposium on applied reconfigurable computing},
  pages={487--498},
  year={2015},
  organization={Springer}
}

@inproceedings{amirany2020retentionlatch,
  title={High-performance and soft error immune spintronic retention latch for highly reliable processors},
  author={Amirany, Abdolah and Jafari, Kian and Moaiyeri, Mohammad Hossein},
  booktitle={2020 28th Iranian Conference on Electrical Engineering (ICEE)},
  pages={1--5},
  year={2020},
  organization={IEEE}
}

@ARTICLE{rajaei2016radiationhardened,
  author={Rajaei, Ramin},
  journal={IEEE Transactions on Magnetics}, 
  title={Radiation-Hardened Design of Nonvolatile MRAM-Based FPGA}, 
  year={2016},
  volume={52},
  number={10},
  pages={1-10},
  doi={10.1109/TMAG.2016.2578278}}

@article{ramezani2020dynamicscheduling,
author = {Ramezani, Reza},
title = {Dynamic scheduling of task graphs in multi-FPGA systems using critical path},
year = {2021},
issue_date = {Jan 2021},
publisher = {Kluwer Academic Publishers},
address = {USA},
volume = {77},
number = {1},
issn = {0920-8542},
url = {https://doi.org/10.1007/s11227-020-03281-3},
doi = {10.1007/s11227-020-03281-3},
journal = {J. Supercomput.},
month = jan,
pages = {597–618},
numpages = {22}
}

@INPROCEEDINGS{scalera1998contextswitching,
  author={Scalera, S.M. and Vazquez, J.R.},
  booktitle={Proceedings. IEEE Symposium on FPGAs for Custom Computing Machines (Cat. No.98TB100251)}, 
  title={The design and implementation of a context switching FPGA}, 
  year={1998},
  volume={},
  number={},
  pages={78-85},
  doi={10.1109/FPGA.1998.707884}}

@article{anderson2020roc,
author = {Anderson, Jeff and Kayraklioglu, Engin and Sun, Shuai and Crandall, Joseph and Alkabani, Yousra and Narayana, Vikram and Sorger, Volker and El-Ghazawi, Tarek},
title = {ROC: A Reconfigurable Optical Computer for Simulating Physical Processes},
year = {2020},
issue_date = {March 2020},
publisher = {Association for Computing Machinery},
address = {New York, NY, USA},
volume = {7},
number = {1},
issn = {2329-4949},
doi = {10.1145/3380944},
journal = {ACM Trans. Parallel Comput.},
month = mar,
articleno = {8},
numpages = {29}
}

@ARTICLE{elghazawi2008promise,
  author={El-Ghazawi, Tarek and El-Araby, Esam and Huang, Miaoqing and Gaj, Kris and Kindratenko, Volodymyr and Buell, Duncan},
  journal={Computer}, 
  title={The Promise of High-Performance Reconfigurable Computing}, 
  year={2008},
  volume={41},
  number={2},
  pages={69-76},
  doi={10.1109/MC.2008.65}}

@ARTICLE{amirany2021taskschedulable,
  author={Amirany, Abdolah and Jafari, Kian and Moaiyeri, Mohammad Hossein},
  journal={IEEE Magnetics Letters}, 
  title={A Task-Schedulable Nonvolatile Spintronic Field-Programmable Gate Array}, 
  year={2021},
  volume={12},
  number={},
  pages={1-4},
  doi={10.1109/LMAG.2021.3092995}}

@article{bakhtiary2023ternarysram,
title = {An SEU-hardened ternary SRAM design based on efficient ternary C-elements using CNTFET technology},
journal = {Microelectronics Reliability},
volume = {140},
pages = {114881},
year = {2023},
issn = {0026-2714},
doi = {https://doi.org/10.1016/j.microrel.2022.114881},
author = {Vahid Bakhtiary and Abdolah Amirany and Mohammad Hossein Moaiyeri and Kian Jafari},
}

@ARTICLE{glorieux2013dff,
  author={Glorieux, Maximilien and Clerc, Sylvain and Gasiot, Gilles and Autran, Jean-Luc and Roche, Philippe},
  journal={IEEE Transactions on Nuclear Science}, 
  title={New D-Flip-Flop Design in 65 nm CMOS for Improved SEU and Low Power Overhead at System Level}, 
  year={2013},
  volume={60},
  number={6},
  pages={4381-4386},
  doi={10.1109/TNS.2013.2284604}}

@article{bahmanabadi2022quaternarymemory,
doi = {10.1149/2162-8777/ac77bb},
url = {https://doi.org/10.1149/2162-8777/ac77bb},
year = {2022},
month = {jun},
publisher = {IOP Publishing},
volume = {11},
number = {6},
pages = {061007},
author = {BahmanAbadi, Motahareh and Amirany, Abdolah and Jafari, Kian and Moaiyeri, Mohammad Hossein},
title = {Efficient and Highly Reliable Spintronic Non-volatile Quaternary Memory Based on Carbon Nanotube FETs and Multi-TMR MTJs},
journal = {ECS Journal of Solid State Science and Technology},
}

@ARTICLE{wang2020nonvolatilesram,
  author={Wang, Chengzhi and Zhang, Deming and Zhang, Kaili and Zeng, Lang and Wang, You and Hou, Zhengyi and Zhang, Youguang and Zhao, Weisheng},
  journal={IEEE Transactions on Electron Devices}, 
  title={Magnetic Nonvolatile SRAM Based on Voltage-Gated Spin-Orbit-Torque Magnetic Tunnel Junctions}, 
  year={2020},
  volume={67},
  number={5},
  pages={1965-1971},
  doi={10.1109/TED.2020.2982683}}

@article{wang2015perpendicular,
doi = {10.1088/0022-3727/48/6/065001},
url = {https://doi.org/10.1088/0022-3727/48/6/065001},
year = {2015},
month = {jan},
publisher = {IOP Publishing},
volume = {48},
number = {6},
pages = {065001},
author = {Wang, Zhaohao and Zhao, Weisheng and Deng, Erya and Klein, Jacques-Olivier and Chappert, Claude},
title = {Perpendicular-anisotropy magnetic tunnel junction switched by spin-Hall-assisted spin-transfer torque},
journal = {Journal of Physics D: Applied Physics}
}

@INPROCEEDINGS{amirany2021mtmrsnqm,
  author={Amirany, Abdolah and Moaiyeri, Mohammad Hossein and Jafari, Kian},
  booktitle={2021 IEEE 51st International Symposium on Multiple-Valued Logic (ISMVL)}, 
  title={MTMR-SNQM: Multi-Tunnel Magnetoresistance Spintronic Non-volatile Quaternary Memory}, 
  year={2021},
  volume={},
  number={},
  pages={172-177},
  doi={10.1109/ISMVL51352.2021.00037}}

@ARTICLE{raouf2023fulladder,
  author={Raouf, Mina and Timarchi, Somayeh},
  journal={IEEE Transactions on Circuits and Systems II: Express Briefs}, 
  title={Non-Volatile and High-Performance Cascadable Spintronic Full-Adder With No Sensitivity to Input Scheduling}, 
  year={2023},
  volume={70},
  number={6},
  pages={2236-2240},
  doi={10.1109/TCSII.2023.3237855}}

@article{grollier2020neuromorphic,
  author = {Julie Grollier and Damien Querlioz and Kerem Camsari and Karin Everschor-Sitte and Shunsuke Fukami and Mark Stiles},
  title = {Neuromorphic Spintronics},
  year = {2020},
  number = {3},
  month = {2020-03-02 00:03:00},
  journal = {Nature Electronics},
  doi = {https://doi.org/10.1038/s41928-019-0360-9},
  language = {en},
}

@ARTICLE{stephan2020spinhall,
  author={Stephan, Andrew W. and Koester, Steven J.},
  journal={IEEE Transactions on Electron Devices}, 
  title={Spin Hall MTJ Devices for Advanced Neuromorphic Functions}, 
  year={2020},
  volume={67},
  number={2},
  pages={487-492},
  doi={10.1109/TED.2019.2959732}}

@ARTICLE{jamshidi2018puremagnetic,
  author={Jamshidi, Vahid and Fazeli, Mahdi},
  journal={IEEE Transactions on Magnetics}, 
  title={Pure Magnetic Logic Circuits: A Reliability Analysis}, 
  year={2018},
  volume={54},
  number={10},
  pages={1-10},
  doi={10.1109/TMAG.2018.2846623}}

@article{ywang2014compactmodel,
title = {Compact model of magnetic tunnel junction with stochastic spin transfer torque switching for reliability analyses},
journal = {Microelectronics Reliability},
volume = {54},
number = {9},
pages = {1774-1778},
year = {2014},
note = {SI: ESREF 2014},
issn = {0026-2714},
doi = {https://doi.org/10.1016/j.microrel.2014.07.019},
author = {Y. Wang and Y. Zhang and E.Y. Deng and J.O. Klein and L.A.B. Naviner and W.S. Zhao},
}

@INPROCEEDINGS{amirany2021stochasticneuron,
  author={Amirany, Abdolah and Meghdadi, Masoud and Moaiyeri, Mohammad Hossein and Jafari, Kian},
  booktitle={2021 26th International Computer Conference, Computer Society of Iran (CSICC)}, 
  title={Stochastic Spintronic Neuron with Application to Image Binarization}, 
  year={2021},
  volume={},
  number={},
  pages={1-5},
  doi={10.1109/CSICC52343.2021.9420559}}

@ARTICLE{zhao2012crosspoint,
  author={Zhao, Weisheng and Chaudhuri, Sumanta and Accoto, Celso and Klein, Jacques-Olivier and Chappert, Claude and Mazoyer, Pascale},
  journal={IEEE Transactions on Nanotechnology}, 
  title={Cross-Point Architecture for Spin-Transfer Torque Magnetic Random Access Memory}, 
  year={2012},
  volume={11},
  number={5},
  pages={907-917},
  doi={10.1109/TNANO.2012.2206051}}

@ARTICLE{rockett1988seulatch,
  author={Rockett, L.R.},
  journal={IEEE Transactions on Nuclear Science}, 
  title={An SEU-hardened CMOS data latch design}, 
  year={1988},
  volume={35},
  number={6},
  pages={1682-1687},
  doi={10.1109/23.25522}}

@ARTICLE{jain2005seureliability,
  author={Jain, P. and Vasi, J. and Lal, R.K.},
  journal={IEEE Transactions on Device and Materials Reliability}, 
  title={SEU reliability analysis of advanced deep-submicron transistors}, 
  year={2005},
  volume={5},
  number={2},
  pages={289-295},
  doi={10.1109/TDMR.2005.848325}}

@ARTICLE{wu2016temperatureimpact,
  author={Wu, Bi and Cheng, Yuanqing and Yang, Jianlei and Todri-Sanial, Aida and Zhao, Weisheng},
  journal={IEEE Transactions on Reliability}, 
  title={Temperature Impact Analysis and Access Reliability Enhancement for 1T1MTJ STT-RAM}, 
  year={2016},
  volume={65},
  number={4},
  pages={1755-1768},
  doi={10.1109/TR.2016.2608910}}

@ARTICLE{amirany2020trng,
  author={Amirany, Abdolah and Jafari, Kian and Moaiyeri, Mohammad Hossein},
  journal={IEEE Transactions on Nanotechnology}, 
  title={True Random Number Generator for Reliable Hardware Security Modules Based on a Neuromorphic Variation-Tolerant Spintronic Structure}, 
  year={2020},
  volume={19},
  number={},
  pages={784-791},
  doi={10.1109/TNANO.2020.3034818}}

@inproceedings{zhou2018rosetta,
author = {Zhou, Yuan and Gupta, Udit and Dai, Steve and Zhao, Ritchie and Srivastava, Nitish and Jin, Hanchen and Featherston, Joseph and Lai, Yi-Hsiang and Liu, Gai and Velasquez, Gustavo Angarita and Wang, Wenping and Zhang, Zhiru},
title = {Rosetta: A Realistic High-Level Synthesis Benchmark Suite for Software Programmable FPGAs},
year = {2018},
isbn = {9781450356145},
publisher = {Association for Computing Machinery},
address = {New York, NY, USA},
url = {https://doi.org/10.1145/3174243.3174255},
doi = {10.1145/3174243.3174255},
booktitle = {Proceedings of the 2018 ACM/SIGDA International Symposium on Field-Programmable Gate Arrays},
pages = {269–278},
numpages = {10},
location = {Monterey, CALIFORNIA, USA},
series = {FPGA '18}
}

@inproceedings{suvizi2025auto,
  title={Auto-Healer: Self-Healing Hardware for Perception Stage Faults in Autonomous Driving Systems},
  author={Suvizi, Ali and Venkataramani, Guru},
  booktitle={Proceedings of the 39th ACM International Conference on Supercomputing},
  pages={1064--1078},
  year={2025}
}

@article{suvizi2025parallel,
  title={A parallel computational approach for energy-efficient hydraulic analysis of water distribution networks using learning automata},
  author={Suvizi, Ali and Ahmadi, Ruhollah and Zamani, Morteza Saheb and Meybodi, Mohammad Reza},
  journal={Sustainable Computing: Informatics and Systems},
  volume={46},
  pages={101135},
  year={2025},
  publisher={Elsevier}
}

@inproceedings{suvizi2024exploring,
  title={Exploring in-memory accelerators and fpgas for latency-sensitive dnn inference on edge servers},
  author={Suvizi, Ali and Subramaniam, Suresh and Lan, Tian and Venkataramani, Guru},
  booktitle={2024 IEEE Cloud Summit},
  pages={1--6},
  year={2024},
  organization={IEEE}
}

@article{adelkhani2025lpcr,
  title={LPCR: Information-Theory Based Low-Power Code Reordering for Serial Links in Network-on-Ch},
  author={Adelkhani, Morteza and Suvizi, Ali and Arzaghi, Farzaneh and Zamani, Sara and Salehi, Mohammad and Shafique, Muhammad},
  journal={IEEE Transactions on Computer-Aided Design of Integrated Circuits and Systems},
  year={2025},
  publisher={IEEE}
}

\end{document}